\documentclass{article}
\usepackage{spconf,amsmath,amssymb,graphicx}
\usepackage{color,cite,balance,url}

\usepackage[ruled,linesnumbered]{algorithm2e}
\DeclareMathOperator*{\argmin}{argmin}

\usepackage{hyperref}

\title{MIMO-SPEECH: End-to-End Multi-Channel Multi-Speaker\\ Speech  Recognition}
\name{Xuankai Chang$^{1,2}$, Wangyou Zhang$^{2}$, Yanmin Qian$^{2}$$^\dagger$, Jonathan Le Roux$^{3}$, Shinji Watanabe$^{1}$$^\dagger$ \thanks{$^\dagger$ Yanmin Qian and Shinji Watanabe are the corresponding authors.}}
\address{
    $^1$Center for Language and Speech Processing, Johns Hopkins University, USA\\
    $^2$SpeechLab, Department of Computer Science and Engineering, Shanghai Jiao Tong University, China\\
    $^3$Mitsubishi Electric Research Laboratories (MERL), USA\\
}
\begin{document}
\ninept
\maketitle
\begin{abstract}

    Recently, the end-to-end approach has proven its efficacy in monaural multi-speaker speech recognition. However, high word error rates (WERs) still prevent these systems from being used in practical applications. On the other hand, the spatial information in multi-channel signals has proven helpful in far-field speech recognition tasks. In this work, we propose a novel neural sequence-to-sequence (seq2seq) architecture, MIMO-Speech, which extends the original seq2seq to deal with multi-channel input and multi-channel output so that it can fully model multi-channel multi-speaker speech separation and recognition. MIMO-Speech is a fully neural end-to-end framework, which is optimized only via an ASR criterion. It is comprised of: 1) a monaural masking network, 2) a multi-source neural beamformer, and 3) a multi-output speech recognition model. With this processing, the input overlapped speech is directly mapped to text sequences.
    We further adopted a curriculum learning strategy, making the best use of the training set to improve the performance. The experiments on the spatialized wsj1-2mix corpus show that our model can achieve more than $60\%$ WER reduction compared to the single-channel system with high quality enhanced signals (SI-SDR = 23.1 dB) obtained by the above separation function.

\end{abstract}
\section{Introduction}
\label{sec:intro}

The cocktail party problem, where the speech of a target speaker is entangled with noise or speech of interfering speakers, has been a challenging problem in speech processing for more than 60 years \cite{Experiment-Cherry1953}. In recent years, there have been many research efforts based on deep learning addressing the multi-speaker speech separation and recognition problems. These works can be categorized into two classes depending on the type of input signals, %
namely single-channel and multi-channel.

In the single-channel multi-speaker speech separation and recognition tasks, several techniques have been proposed, achieving significant progress. One such technique is deep clustering (DPCL) \cite{DeepClustering-Hershey2015,DeepClustering2-Isik2016,Analysis-Menne2019}. In DPCL, a neural network is trained to map each time-frequency unit to an embedding vector, which is used to assign each unit to a source by a clustering algorithm afterwards. DPCL was then integrated into a joint training framework with end-to-end speech recognition in \cite{End-Settle2018}, showing promising performance. Another approach called permutation-free training \cite{DeepClustering-Hershey2015,DeepClustering2-Isik2016} or permutation-invariant training (PIT) \cite{Permutation-Yu2017,Multitalker-Kolbaek2017} relies on training a neural network to estimate a mask for every speaker with a permutation-free objective function that minimizes the reconstruction loss. %
PIT was later applied to multi-speaker automatic speech recognition (ASR) by directly optimizing a speech recognition loss \cite{Recognizing-Yu2017,Single-Qian2018} within a DNN-HMM hybrid ASR framework. In recent years, end-to-end models have drawn a lot of attention in single-speaker ASR systems and shown great success \cite{Towards-Graves2014,Listen-Chan2016,Joint-Kim2017,Joint-Hori2017}. These models have simplified the ASR paradigm by unifying acoustic, language, and phonetic models into a single neural network. In \cite{End2End-Seki2018,End-Chang2019}, joint CTC/attention-based encoder-decoder \cite{Joint-Kim2017} end-to-end models were developed to solve the single-channel multi-speaker speech recognition problem, where the encoder separates the mixed speech features and the attention-based decoder generates the output sequences.
Although significant performance improvements have been achieved in the monaural case, there is still a large performance gap compared with that of single-speaker speech recognition systems, making such models not yet ready for widespread application in real scenarios. %

The other important case is that of multi-channel multi-speaker speech separation and recognition, where the input signals are collected by microphone arrays. Acquiring multi-channel data is not so limiting nowadays, where microphone arrays are widely deployed in many devices. When multi-channel data is available, the spatial information can be exploited to determine the speaker location and to separate the speech with higher accuracy. Yoshioka et al \cite{Recognizing-Yoshioka2018} proposed a method for performing multi-channel speech separation under the PIT framework. A mask-based beamformer called the unmixing transducer was used to separate the overlapped speech. Another method proposed by Wang et al \cite{Multichannel-Wang2018} leverages the inter-channel differences as spatial features combined with the single-channel spectral features as the input, to separate the multi-channel data using the DPCL technique.

Previous works based on multi-channel multi-speaker input mainly focus on separation. In this paper, we propose an end-to-end multi-channel multi-speaker speech recognition system. Such a sequence-to-sequence model is trained to directly map multi-channel input (MI) speech signals where multiple speakers speak simultaneously, to multiple output (MO) text sequences, one for each speaker. We refer to this system as MIMO-Speech. The recent research on single-speaker far-field speech recognition has shown that neural beamforming techniques for denoising
\cite{Neural-Heymann2016,Improved-Erdogan2016} can achieve state-of-the art results in robust ASR tasks \cite{RWTH-Menne2016,Beamnet-Heymann2017,Frequency-Wu2019}. Several works have shown that it is feasible to design a totally differentiable end-to-end model by integrating the neural beamforming mechanism and the sequence-to-sequence speech recognition together \cite{Multichannel-Ochiai2017,Multi-Braun2018,Stream-Wang2019,Investigation-ShanmugamSubramanian2019}. \cite{Does-Ochiai2017} further shows that the neural beamforming function in a multi-channel end-to-end system can enhance the signals. In light of this success, we redesigned the neural-beamformer front-end to allow it to attend to multiple beams at different directions. After getting the separated signals, the log filter bank features are extracted inside the neural network. Finally, a joint CTC/attention-based encoder-decoder recognizes each feature stream. With this framework, the outputs of the beamformer in the middle of the model can also be used as speech separation signals. During the training, a data scheduling strategy using curriculum learning is specially designed and leads to an additional performance boost. To prove the basic concept of our method, we first evaluated our proposed method in the anechoic scenario. From the results, we find that even without explicitly optimizing for separation, the intermediate signals after the beamformer still show very good quality in terms of audibility. Then we also tested the model on the reverberant case to give a preliminary result.

\section{End-to-End Multi-channel Multi-speaker ASR}
\label{sec:algo}

\begin{figure}[htb]

\begin{minipage}[b]{1.0\linewidth}
  \centering
  \label{fig:model}
  \centerline{\includegraphics[width=\linewidth]{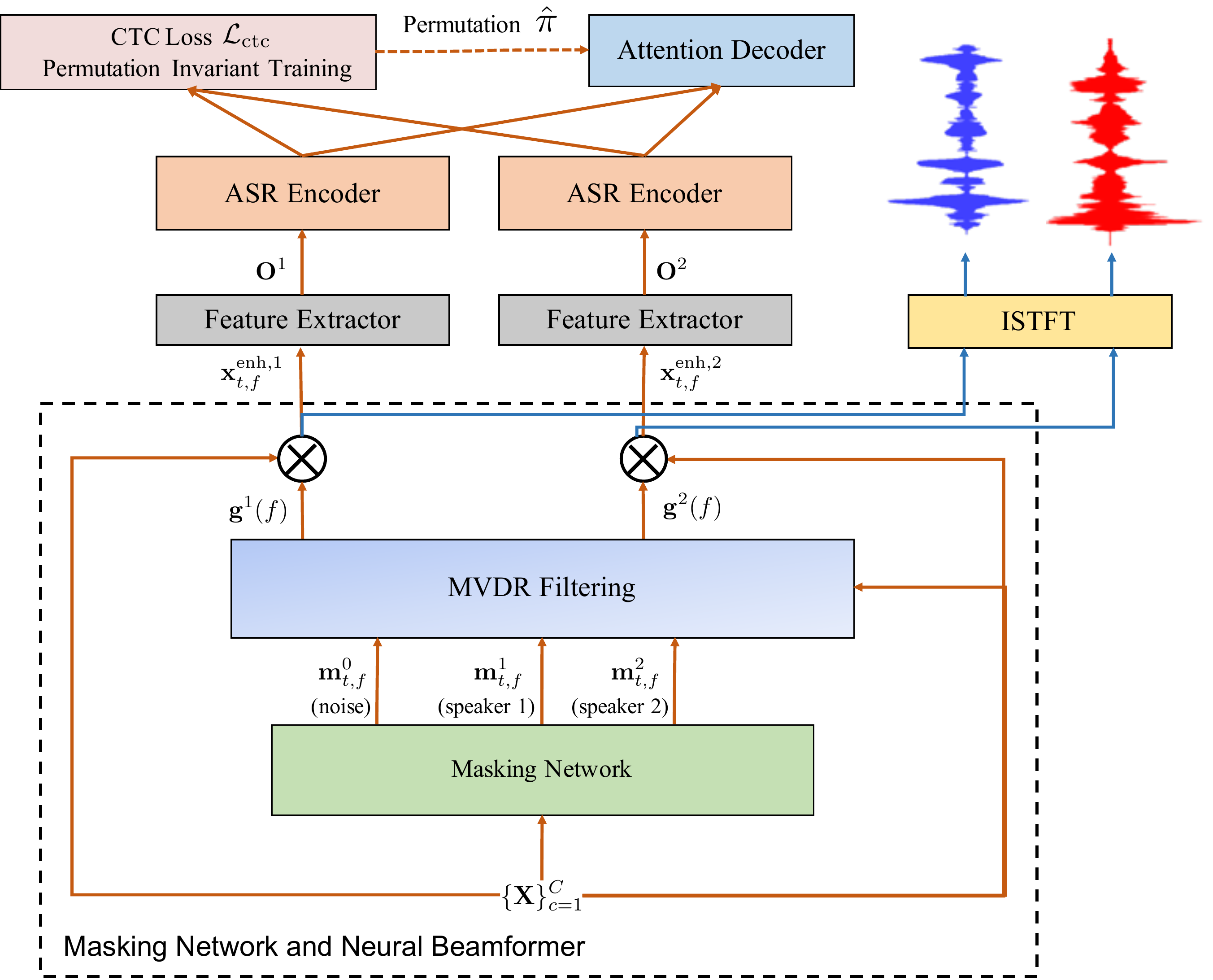}}
\end{minipage}

\caption{End-to-End Multi-channel Multi-speaker Model}
\label{fig:res1}
\end{figure}

In this section, we first present the proposed end-to-end multi-channel multi-speaker speech recognition model, which is shown in Fig.~\ref{fig:res1}. We then describe the techniques applied in scheduling the training data, which have an important role in improving the performance.

\subsection{Model Architecture}
\label{ssec:arch}
By using the differences in the signals recorded at each sensor, distributed sensors can exploit spatial information. %
They are thus particularly useful for separating sources that are spatially partitioned. In this work, we present a sequence-to-sequence architecture with multi-channel input and multi-channel output to model the multi-channel multi-speaker speech recognition, shown in Fig.~\ref{fig:res1} for the case of two speakers. The proposed end-to-end multi-channel multi-speaker ASR model can be divided into three stages. The first stage is a single-channel masking network to perform pre-separation by predicting multiple speaker and noise masks for each channel. Then a multi-source neural beamformer is used to spatially separate multiple speaker sources. In the last stage, an end-to-end ASR module with permutation-free training is used to perform the multi-output speech recognition.

We used a similar architecture as in \cite{Multichannel-Ochiai2017}, where the masking network and the neural beamformer are integrated into an attention-based encoder-decoder neural network, and the whole model is jointly optimized solely via a speech recognition objective. The input of the model can consist of an arbitrary number of channels $C$, and its output is the text sequence for each speaker directly. We denote by $J$ the number of speakers in the mixed utterances, and for simplicity of notation, we shall consider the noise component as the $0$-th source.  %

\subsubsection{Monaural Masking Network}
\label{sssec:masknet}
The monaural masking network, shown at the bottom of Fig.~\ref{fig:res1}, estimates the masks of each channel for every speaker and an extra noise component. Let us denote by $\mathbf{X}_c =(x_{t,f,c})_{t,f} \in \mathbb{C}^{T \times F}$ the complex STFT of the $c$-th channel of the observed multi-channel multi-speaker speech, where $1 \leq t \leq T,\, 1 \leq f \leq F,\, 1 \leq c \leq C$ denote time, frequency, and channel indices, respectively.
The mask estimation module produces time-frequency masks  $\mathbf{M}_{c}^{i}=(m^{i}_{t,f,c})_{t,f} \in [0,1]^{T \times F}$, with $i \in \{1,\dots,J\}$ for each of the $J$ speakers, and $i=0$ for the noise, using the complex STFT of the $c$-th channel of the observed multi-channel multi-speaker speech as input. %
The computation is performed independently on each of the input channels:
\begin{align}
    \mathbf{M}_c = \mathrm{MaskNet} (\mathbf{X}_c),
\end{align}
where $\mathbf{M}_{c} = (\mathbf{M}_{c}^{i})_i \in [0,1]^{T \times F \times J}$ is the set of estimated masks for the $c$-th channel.

\subsubsection{Multi-source Neural Beamformer}
\label{sssec:neuralbeam}
The multi-source neural beamformer is a key component in the proposed model, which produces the separated speech of each speaker. The masks obtained on each channel for each speaker and the noise are used in the computation of the power spectral density (PSD) matrices of each source as follows \cite{NTT-Yoshioka2015,Neural-Heymann2016}: %
\begin{equation}
        \mathbf{\Phi}^{i} (f) = \frac{1}{\sum_{t=1}^T \mathbf{m}^i_{t,f}} \sum_{t=1}^T \mathbf{m}^i_{t,f} \mathbf{x}_{t,f} \mathbf{x}^{H}_{t,f} \; \in \mathbb{C}^{C \times C}, 
\end{equation}
where $i \in \{0,\dots,J\}$,  $\mathbf{x}_{t,f} = \{x_{t,f,c}\}_{c=1}^C$, $\mathbf{m}_{t,f}^{i} = \{m^i_{t,f,c}\}_{c=1}^C$, and $^H$ represents the conjugate transpose.

After getting the PSD matrices of every speaker and the noise, we estimate the beamformer's time-invariant filter coefficients $\mathbf{g}^{i}(f)$ at frequency $f$ for each speaker $i\in\{1,\cdots,J\}$ via the MVDR formalization \cite{Optimal-Souden2009} as follows:
\begin{align}
    \mathbf{g}^{i}(f) &= \frac{(\sum _{j \neq i} \mathbf{\Phi}^{j} (f))^{-1} \mathbf{\Phi}^{i}(f)}{\text{Tr}((\sum _{j \neq i} \mathbf{\Phi}^{j} (f))^{-1} \mathbf{\Phi}^{i}(f))} \mathbf{u} \; \in \mathbb{C}^C, \label{eq:mvdr}
\end{align}
where $\mathbf{u} \in \mathbb{R}^C$ is a vector representing the reference microphone that is derived from an attention mechanism \cite{Multichannel-Ochiai2017}, and $\text{Tr}(\cdot)$ denotes the trace operation. Notice that in Eq.~\ref{eq:mvdr}, the formula to derive the filter coefficient is different from that in \cite{Multichannel-Ochiai2017} in the way that the noise PSD is replaced by $\sum_{ j \neq i } \mathbf{\Phi}^{j}(f)$. This is because both noise and other speakers are considered as interference when focusing on a given speaker. %
This is akin to the 
speech-speech-noise (SSN) model in \cite{Recognizing-Yoshioka2018}. Such a method is employed to make more accurate estimations of the PSD matrices, in which the traditional PSD matrix is expressed using 
the PSD matrix of interfering speaker and that of the background noise.

Finally, the beamforming filters $\mathbf{g}^{i}(f)$ obtained in Eq.~\ref{eq:mvdr} are used to separate and denoise the input overlapped multi-channel signals $\mathbf{x}_{t,f} \in \mathbb{C}^{C}$ to obtain a single-channel estimate of the enhanced STFT $\hat{s}^{i}_{t,f}$ for speaker $i$:
\begin{align}
    \hat{s}^{i}_{t,f} = (\mathbf{g}^{i}(f))^{H} \mathbf{x}_{t,f} &  \in \mathbb{C}.
\end{align}
Each separated speech signal waveform can be obtained by inverse STFT for listening, as $\mathrm{iSTFT}(\hat{\mathbf{S}}^i), \; i=1,\dots,J$. %

\subsubsection{End-to-End Speech Recognition}
\label{sssec:e2e-asr}
The outputs of the neural beamformer are estimates of the separated speech signals for each speaker. Before feeding these streams to the end-to-end speech recognition submodule, we need to convert the STFT features to normalized log filterbank features. %
A log mel filterbank transformation is first applied on the magnitude of the beamformed STFT signal $\hat{\mathbf{S}}^{i}=(\hat{S}^{i}_{t,f})_{t,f}$ for each speaker $i$, and a global mean-variance normalization is then performed on the log-filterbank feature to produce a proper input $\mathbf{O}^i$ for the speech recognition submodule:
\begin{align}
    \mathbf{FBank}^i &= \mathrm{MelFilterBank}(|\hat{\mathbf{S}}^{i}|), \\
    \mathbf{O}^i &= \mathrm{GlobalMVN}(\log(\mathbf{FBank}^i)).
\end{align}

We briefly introduce the end-to-end speech recognition submodule used here, which is similar to the joint CTC/attention-based encoder-decoder architecture \cite{Joint-Kim2017}. The feature vectors $\mathbf{O}^i$ are first transformed to a hidden representation $\mathbf{H}^i$ by an encoder network. A decoder then generates the output token sequences based on the history information $\mathbf{y}$ and a weighted sum vector $\mathbf{c}$ obtained with an attention mechanism. The end-to-end speech recognition is computed as follows:
\begin{align}
    \mathbf{H}^i &= \mathrm{Encoder} (\mathbf{O}^i) \\
    \mathbf{c}^i_n, \mathbf{\alpha}^i_n &= \mathrm{Attention} (\mathbf{\alpha}^i_{n-1}, \mathbf{e}^i_{n-1}, \mathbf{H}^i) \\
    \mathbf{e}^i_{n} &= \mathrm{Update}(\mathbf{e}^i_{n-1}, \mathbf{c}^i_{n-1}, \mathbf{y}^i_{n-1}) \\
    \mathbf{y}^i_n &\sim \mathrm{Decoder} (\mathbf{c}^i_n, \mathbf{y}^i_{n-1}),
\end{align}
where $i$ denotes the index of the source stream and $n$ an output label sequence index.

Typically, the history information $\mathbf{y}$ is replaced by the reference labels $\mathbf{R} = (r_1, \cdots, r_N)$ in a teacher-forcing fashion at training time. However, since there are multiple possible assignments between the inputs and the references, it is necessary to used permutation invariant training (PIT) in the end-to-end speech recognition \cite{End2End-Seki2018,End-Chang2019}. The best permutation of the input sequences and the references is determined by the connectionist temporal classification (CTC) loss $\mathrm{Loss}_{\text{ctc}}$:
\begin{align}
    \hat{\pi} = \argmin_{\pi \in \mathcal P} \sum_i \mathrm{Loss}_{\text{ctc}} (\mathbf{Z}^i, \mathbf{R}^{\pi(i)}),\; i=1,\dots,J,
\end{align}
where $\mathbf{Z}^i$ denotes the output sequence computed from the encoder output $\mathbf{H}^i$ for the CTC loss, $\mathcal P$ is the set of all permutations on $\{1,\dots,J\}$, and $\pi(i)$ is the $i$-th element for permutation $\pi$.

The final ASR loss of the model is obtained as:
\begin{align}
    \mathcal{L} &= \lambda \mathcal{L}_{\text{ctc}} + (1-\lambda) \mathcal{L}_{\text{att}}, \label{eq:loss} \\
    \mathcal{L}_{\text{ctc}} &= \sum_i \mathrm{Loss}_{\text{ctc}} (\mathbf{Z}^i, \mathbf{R}^{\hat{\pi}(i)}), \\
    \mathcal{L}_{\text{att}} &= \sum_i \mathrm{Loss}_{\text{att}} (\mathbf{Y}^i, \mathbf{R}^{\hat{\pi}(i)}),
\end{align}
where $0 \leq \lambda \leq 1$ is an interpolation factor, and $\mathrm{Loss}_{\text{att}}$ is the cross-entropy loss to train the attention-based encoder-decoder networks.

\subsection{Data Scheduling and Curriculum Learning}
\label{ssec:data_schedule}

From preliminary empirical results, we find that it is relatively difficult to perform straightforward end-to-end training of such a multi-stage model, especially without an intermediate criterion to guide the training. In our model, the speech recognition submodule has the same architecture as the typical end-to-end speech recognition model, and the input is expected to be similar to the log filterbank of single-speaker speech. Thus, in order to train the model properly, we did not only use the spatialized utterances of the multi-speaker corpus but also the single-speaker utterances from the original WSJ training set. During training, every batch is randomly chosen either from the multi-channel multi-speaker set or from the single-channel single-speaker set. For single-speaker batches, the masking network and neural beamformer stages are bypassed, and the input is directly fed to the recognition submodule. Furthermore, the loss is calculated without considering permutations, as there is only a single speaker per input.

With this data scheduling scheme, the model can achieve a decent performance from random initialization. %
For multi-channel multi-speaker data batches, the loss of the ASR objective function is back-propagated down through the model to the masking network. For data batches consisting of single-speaker utterances, only the speech recognition part is optimized, which leads to more accurate loss computation in the future. The single-speaker data batches rectify the behavior of the ASR model as it performs regularization during the training.

According to previous researches, starting from easier subtasks can lead the model to learn better, an approach called curriculum learning \cite{Curriculum-Bengio2009,Deep-Amodei2016}. To further exploit the data scheduling scheme, we introduce more constraints on the order of the data batches of the training set. As was observed in prior research by \cite{Single-Qian2018}, the signal-to-noise ratio (SNR, the energy ratio between the target speech and the interfering sources) has a great influence on the final recognition performance. When the speech energy levels of the target speaker and the interfering sources are obviously different, the recognition accuracy of the interfering source speech is very poor. Thus, we sort the multi-speaker data in ascending order of SNR between the loudest and quietest speaker, thus starting with mixtures where both speakers are at similar levels. Furthermore, we sort the single-speaker data from short to long, as short sequences tend to be easier to learn in seq2seq learning. The strategy is formally depicted in Algorithm~\ref{algo:cl}. We applied such a curriculum learning strategy in order to make the model learn step by step and expect it to improve the training.

\begin{algorithm}
    \SetAlgoLined
     Load the training dataset $\mathbf{X}$\;
     Categorize the training data $\mathbf{X}$ into single-channel single-speaker data $\mathbf{X}_{\text{clean}}$ and multi-channel multi-speaker data $\mathbf{X}_{\text{noisy}}$\;
     Sort the single-channel single-speaker training data in $\mathbf{X}_{\text{clean}}$ in ascending order of the utterance lengths, leading to $\mathbf{X}'_{\text{clean}}$\;
     Sort the multi-channel multi-speaker training data in $\mathbf{X}_{\text{noisy}}$ in ascending order of the SNR level, leading to $\mathbf{X}'_{\text{noisy}}$ \;
     Divide $\mathbf{X}'_{\text{clean}}$ and $\mathbf{X}'_{\text{noisy}}$ into minibatch sets $\mathbf{\mathcal{B}}_{\text{clean}}$ and $\mathbf{\mathcal{B}}_{\text{noisy}}$\;
     Sort batches to alternate between batches from $\mathbf{\mathcal{B}}_{\text{clean}}$ and $\mathbf{\mathcal{B}}_{\text{noisy}}$\;
     \While{model is not converged}{
       \For{each $b$ in all minibatches}{
          Feed minibatch $b$ into the model, update the model\;
       }
     }
     \While{model is not converged}{
       Shuffle the training data in $\mathbf{X}_{\text{clean}}$ and $\mathbf{X}_{\text{noisy}}$  randomly and divide them into minibatch sets $\mathbf{\mathcal{B}}'_{\text{clean}}$ and $\mathbf{\mathcal{B}}'_{\text{noisy}}$\; \label{algostep:shuf1}
       Select each minibatch randomly from $\mathbf{\mathcal{B}}'_{\text{clean}}$ and $\mathbf{\mathcal{B}}'_{\text{noisy}}$ and feed it into the model iteratively to update the model\; \label{algostep:shuf2}
     }
     \caption{Curriculum learning strategy}
    \label{algo:cl}
\end{algorithm} 
\section{Experiment}
\label{sec:experiment}
To check the effectiveness of our proposed end-to-end model, we evaluated it on the remixed WSJ1 data used in \cite{End2End-Seki2018}, which we here refer to as the wsj1-2mix dataset. The multi-speaker speech training set was generated by randomly selecting two utterances from the WSJ SI284 corpus, resulting in a $98.5$ h dataset. The signal-to-noise ratio (SNR) of one source against the other was randomly chosen from a uniform distribution in the range of  $[-5, 5]$ dB. The validation and evaluation sets were generated in a similar way by selecting source utterances from the WSJ Dev93 and Eval92 respectively, and the durations are $1.3$ h and $0.8$ h. We then create a new spatialized version of the wsj1-2mix dataset following the process applied to the wsj0-2mix dataset in \cite{Multichannel-Wang2018}, using a room impulse response (RIR) generator\footnote{Available online at \url{https://github.com/ehabets/RIR-Generator}}, where the characteristics of each two-speaker mixture are randomly generated including room dimensions, speaker locations, and microphone geometry\footnote{The spatialization toolkit is available at \url{http://www.merl.com/demos/deep-clustering/spatialize_wsj0-mix.zip}}.

To train the model, we used the spatialized wsj1-2mix data with $J=2$ speakers as well as the train\_si284 training set from the WSJ1 dataset to regularize the training procedure. All input data are raw waveform audio signals. 
The STFT was computed using $25$ ms-width Hann window with $10$ ms shift, with zero-padding resulting in a spectral dimension $F=257$. 
In our experiments, we only report results in the case of $C=2$ channels, but our model is flexible and can be used with an arbitrary number of channels. We first report recognition and separation results in the anechoic scenario in Sections~\ref{ssec:result-recognition} and \ref{ssec:result-separation}. Then we show  preliminary results in the reverberant scenario in Section~\ref{ssec:result-reverb}.

\subsection{Configurations}
\label{ssec:config}
Our end-to-end multi-channel multi-speaker model is completely built based on the ESPnet framework \cite{ESPnet-Watanabe2018} with Pytorch backend. All the network parameters were initialized randomly from uniform distribution in the range $[-0.1, 0.1]$. We used AdaDelta with $\rho = 0.95$ and $\epsilon=1e^{-8}$ as optimization method. The maximum number of epochs for training is set to $15$ but the training process is stopped early if performance does not increase for 3 consecutive epochs. For decoding, a word-based language model \cite{End-Hori2018} was trained on the transcripts of the WSJ corpus.

\subsubsection{Neural Beamformer}
\label{sssec:beamform}
The mask estimation network is a 3-layer bidirectional long-short term memory with projection (BLSTMP) network with 512 cells in each direction. The computation of the reference microphone vector has the same parameters as in \cite{Multichannel-Ochiai2017} except the vector dimension which is here set to $512$. In the MVDR formula of Eq.~\ref{eq:mvdr}, a small value $\epsilon$ is added to the PSD matrix to guarantee that an inverse exists. %

\subsubsection{Encoder-Decoder Network}
\label{sssec:enc-dec}
The encoder network consists of two VGG-motivated CNN blocks and three BLSTMP layers. The CNN layers have a kernel size of $3\times3$ and the number of feature maps is $64$ and $128$ in the first and second block, respectively. Every BLSTMP layer has 1024 memory cells in each direction with projection size 1024. $80$ dimensional log filterbank features are extracted for each separated speech signals and global mean-variance normalization is applied, using the statistics of the single-speaker WSJ1 training set. In the decoder network, there is only a single layer of unidirectional long-short term memory network (LSTM) and the number of cells is 300. The interpolation factor $\lambda$ of the loss function in Eq.~\ref{eq:loss} is set to $0.2$.

\subsection{Performance of Multi-Speaker Speech Recognition}
\label{ssec:result-recognition}
In this subsection, we describe the speech recognition performance on the spatialized anechoic wsj1-2mix data, which only modifies the signals via  delays and decays due to the propagation. Note that although beamforming algorithms can address the anechoic case without too much effort, it is still necessary to show that our proposed end-to-end method can address the multi-channel multi-speaker speech recognition problem and both the speech recognition submodule and the neural beamforming separation submodule perform well as they are designed. We shall also note that the whole system is trained solely through an ASR objective, and it is thus not trivial for the system to learn how to properly separate the signals even in the anechoic case.

The multi-speaker speech recognition performance is shown in Table~\ref{tab:results-asr}. There are three single-channel end-to-end speech recognition baseline systems.
The first one is a single-channel multi-speaker ASR model trained on the first channel of the spatialized corpus, where the model is the same as the one proposed in \cite{End-Chang2019}. 
The second is a single-channel multi-speaker ASR model trained with speech that is enhanced by BeamformIt \cite{Acoustic-Anguera2007}, which is a well-known delay-and-sum beamformer.
And the third one is to use BeamformIt to first separate the speech by choosing its best and second-best output streams, and then to recognize them with a normal single-speaker end-to-end ASR model.
The spatialization of the corpus results in a degradation of the performance: the multi-speaker ASR model trained with the 1st channel has a word error rate (WER) of $29.43\%$ on the evaluation set, compared to only $20.43\%$ obtained on the original unspatialized wsj1-2mix data in \cite{End-Chang2019}. Using the BeamformIt tool to enhance the spatialized signal can improve the recognition accuracy of a multi-speaker model, leading to a WER of $21.75\%$ on the evaluation set. However, traditional beamforming algorithms such as BeamformIt can not perfectly separate the overlapped speech signals, and the performance of the single-speaker model in terms of WER is very poor, $98.00\%$.

The performance of our proposed end-to-end multi-channel multi-speaker model (MIMO-Speech) is shown at the bottom of the table. The curriculum learning strategy described in Section~\ref{ssec:data_schedule} is used to further improve performance. From the table, it can be observed that MIMO-Speech is significantly better than traditional methods, achieving $4.51\%$ character error rate (CER) and $8.62\%$ word error rate (WER). Compared against the best baseline model, the relative improvement is over $60\%$ in terms of both CER and WER. When applying our data scheduling scheme by sorting the multi-speaker speech in ascending order of SNRs, an additional performance boost can be realized. The final CER and WER on the evaluation set are $3.75\%$ and $7.55\%$ respectively, with over $12\%$ relative improvement against MIMO-Speech without curriculum learning. Overall, our proposed MIMO-Speech network can achieve good recognition performance on the spatialized anechoic wsj1-2mix corpus.

\begin{table}[th]
    \caption{Performance in terms of average CER and WER [\%] on the spatialized anechoic wsj1-2mix corpus.}
    \label{tab:results-asr}
    \vspace{0.1cm}
    \centering
    \small
    \begin{tabular}{l | c | c}
        \hline \hline
        Model & \!dev CER\! & \!eval CER\! \\
        \hline
        2-spkr ASR (1st channel) & 22.65 & 19.07 \\
        BeamformIt Enhancement (2-spkr ASR) \!\!& 15.23 & 12.45 \\
        BeamformIt Separation (1-spkr ASR) & 77.30 & 77.10 \\
        \hline
        MIMO-Speech & \phantom{1}7.29 & \phantom{1}4.51 \\
        \, + Curriculum Learning (SNRs) & \textbf{\phantom{1}6.34} & \textbf{\phantom{1}3.75} \\
        \hline \hline
        Model & \!\!dev WER\!\! & \!\!eval WER\!\! \\
        \hline
        2-spkr ASR (1st channel) & 34.98 & 29.43 \\
        BeamformIt Enhancement (2-spkr ASR) \!\!\!& 26.61 & 21.75 \\
        BeamformIt Separation (1-spkr ASR) & 98.60 & 98.00 \\
        \hline
        MIMO-Speech & 13.54 & \phantom{1}8.62 \\
        \, + Curriculum Learning (SNRs) & \textbf{12.59} & \textbf{\phantom{1}7.55} \\
        \hline \hline
    \end{tabular}
\end{table}

\subsection{Performance of Multi-Speaker Speech Separation}
\label{ssec:result-separation}
\begin{figure}[htb]
  \vspace{-.3cm}
\begin{minipage}[b]{0.46\linewidth}
  \centering
  \centerline{\includegraphics[width=1.2\linewidth]{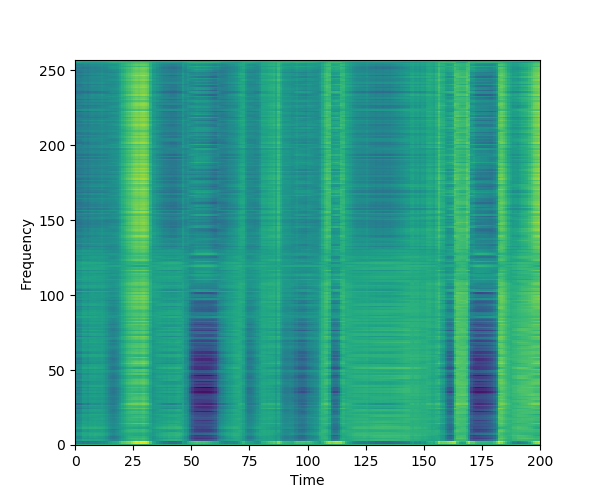}}
  \centerline{(a) Mask for Speaker 1}\medskip
\end{minipage}
\begin{minipage}[b]{0.46\linewidth}
  \centering
  \centerline{\includegraphics[width=1.2\linewidth]{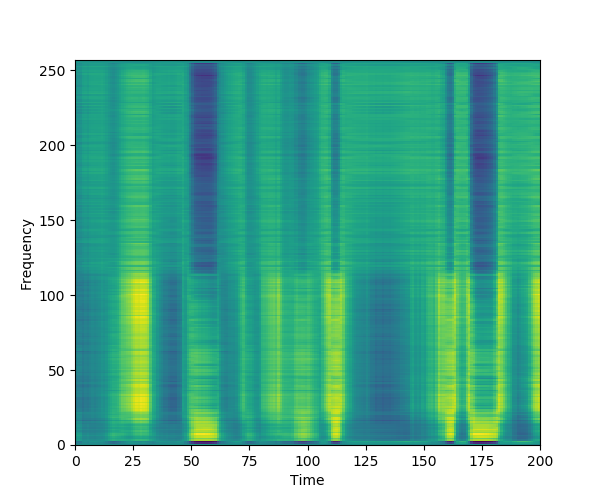}}
  \centerline{(b) Mask for Speaker 2}\medskip
\end{minipage}
\hfill
\begin{minipage}[b]{0.46\linewidth}
  \centering
  \centerline{\includegraphics[width=1.2\linewidth]{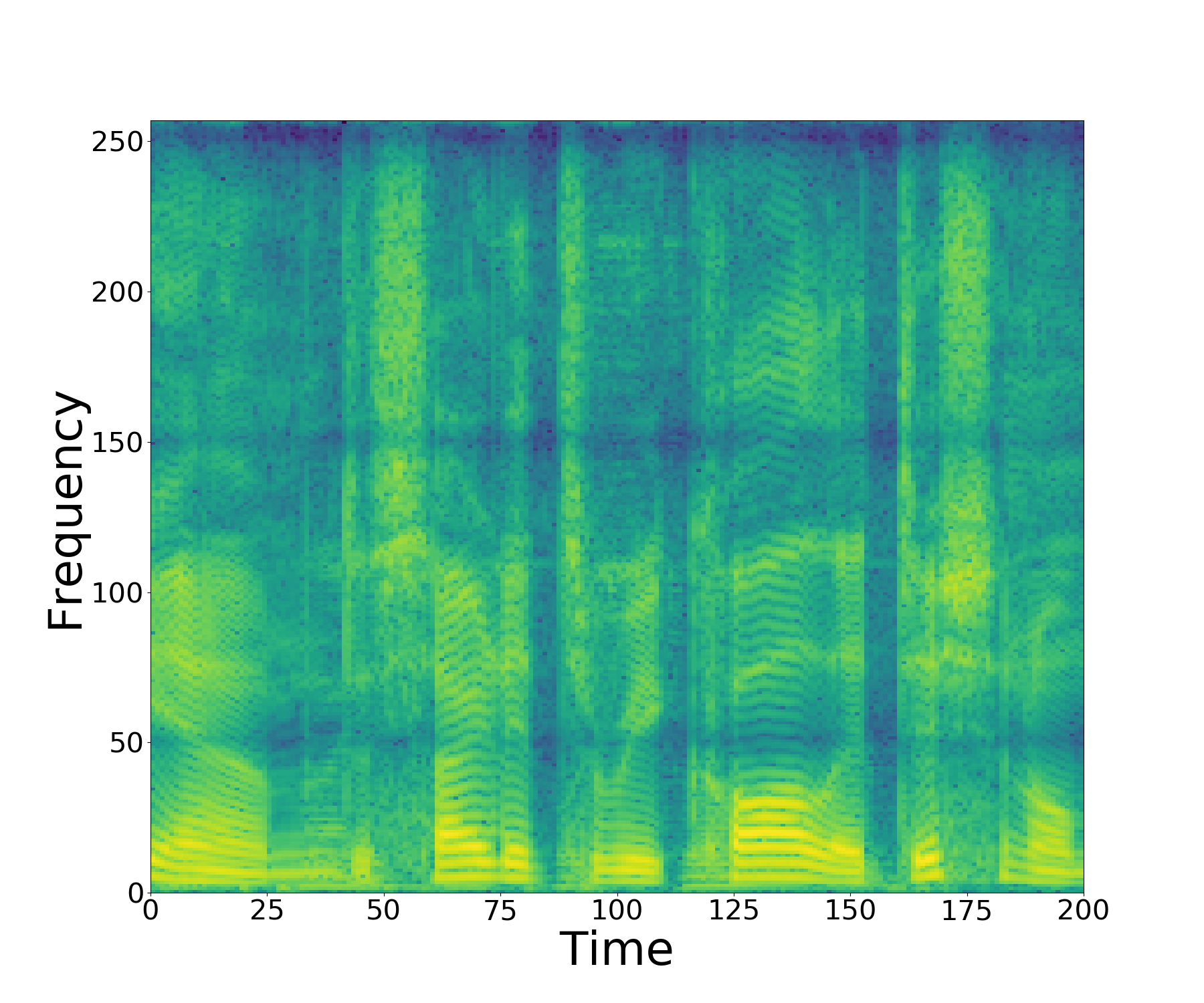}}
  \centerline{(c) Separated Speech for Speaker 1}\medskip
\end{minipage}
\begin{minipage}[b]{0.46\linewidth}
  \centering
  \centerline{\includegraphics[width=1.2\linewidth]{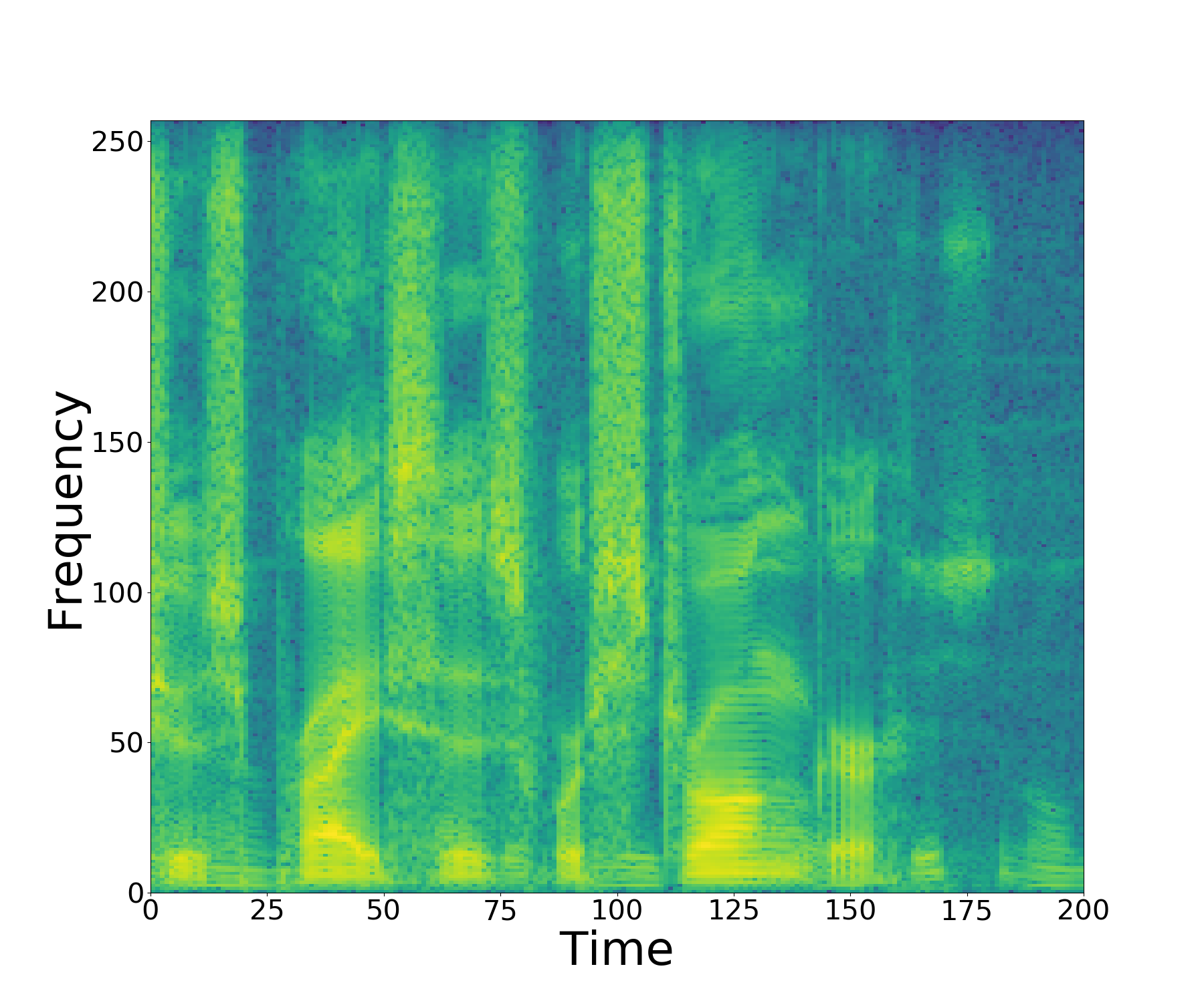}}
  \centerline{(d) Separated Speech for Speaker 2}\medskip
\end{minipage}
\hfill
\begin{minipage}[b]{\linewidth}
  \centering
  \centerline{\includegraphics[width=0.55\linewidth]{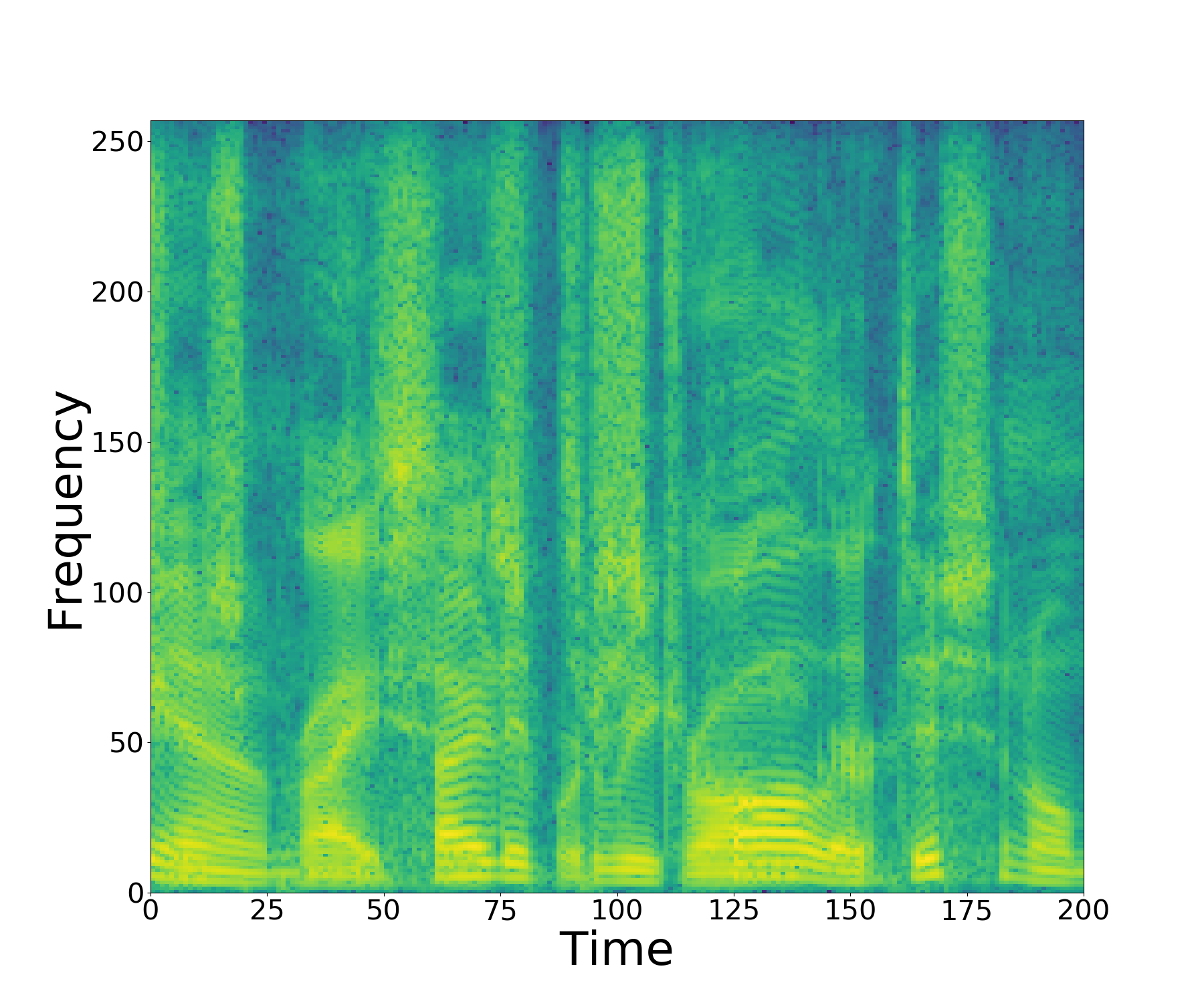}}
  \centerline{(e) Overlapped Speech}\medskip
\end{minipage}
\caption{Example of masks output by the masking network and separated speech log spectrograms output by the MVDR beamformer.}
\label{fig:spec}
\end{figure}

One question regarding our model is whether the front-end of MIMO-Speech, the neural beamformer, learns a proper beamforming behavior as other algorithms do since there is no explicit speech separation criterion to optimize the network. To investigate the role of the neural beamformer, we consider the masks $\mathbf{m}^{i}$ that are used to compute the PSD matrices and the enhanced separated STFT signals $\hat{\mathbf{s}}^{i}, i=1,\dots,J$ obtained at the output of the beamformer. Example results are shown in Fig.~\ref{fig:spec}. Note that in our model, the masks are not constrained to sum to 1 at each time-frequency unit, resulting in a scaling indeterminacy within each frequency. For better readability in the figures, we here renormalize each mask using its median within each frequency. In the figure, the difference between the masks from each speaker is clear. And from the spectrogram, it is also observed that for each separated stream, the signals are less overlapped compared with the input multi-speaker speech signal. The mask and spectrogram examples suggest that MIMO-Speech can separate the speech to some level.

To evaluate the separation quality, we reconstruct the separated waveforms for each speaker from the outputs of the beamformer via inverse STFT, and compare them with the reference signals in terms of PESQ and scale-invariant signal-to-distortion ratio (SI-SDR) \cite{LeRoux2019SISDR}. The results are shown in Table~\ref{tab:results-separation}. As we can see, the separated audios have very good quality. The separated signals from the MIMO-Speech model \footnote{Audio samples are available online at \url{https://simpleoier.github.io/MIMO-Speech/index.html}} have an average PESQ value of $3.6$ and an average SI-SDR of $23.1$ dB. When using curriculum learning, PESQ and SI-SDR degrade slightly, but the quality is still very high. This result suggests that our proposed MIMO-Speech model is capable of learning to separate overlapped speech via beamforming, based solely on an ASR objective.

\begin{table}[th]
    \caption{Performance in terms of average PESQ and SI-SDR [dB] on the spatialized anechoic wsj1-2mix corpus.}
    \label{tab:results-separation}
    \vspace{0.1cm}
    \centering
    \small
    \begin{tabular}{l | c | c}
        \hline \hline
        Model & dev PESQ & eval PESQ \\
        \hline
        MIMO-Speech & \phantom{1}3.6 & \phantom{1}3.6 \\
        \, + Curriculum Learning (SNRs) & \textbf{\phantom{1}3.7} & \textbf{\phantom{1}3.6} \\
        \hline \hline
        Model & dev SI-SDR & eval SI-SDR \\
        \hline
        MIMO-Speech & \textbf{22.1} & \textbf{23.1} \\
        \, + Curriculum Learning (SNRs) & 21.1 & 21.8 \\
        \hline \hline
    \end{tabular}
\end{table}

\begin{figure}[htb]
\begin{minipage}[b]{1.0\linewidth}
  \centering
  \centerline{\includegraphics[width=\linewidth]{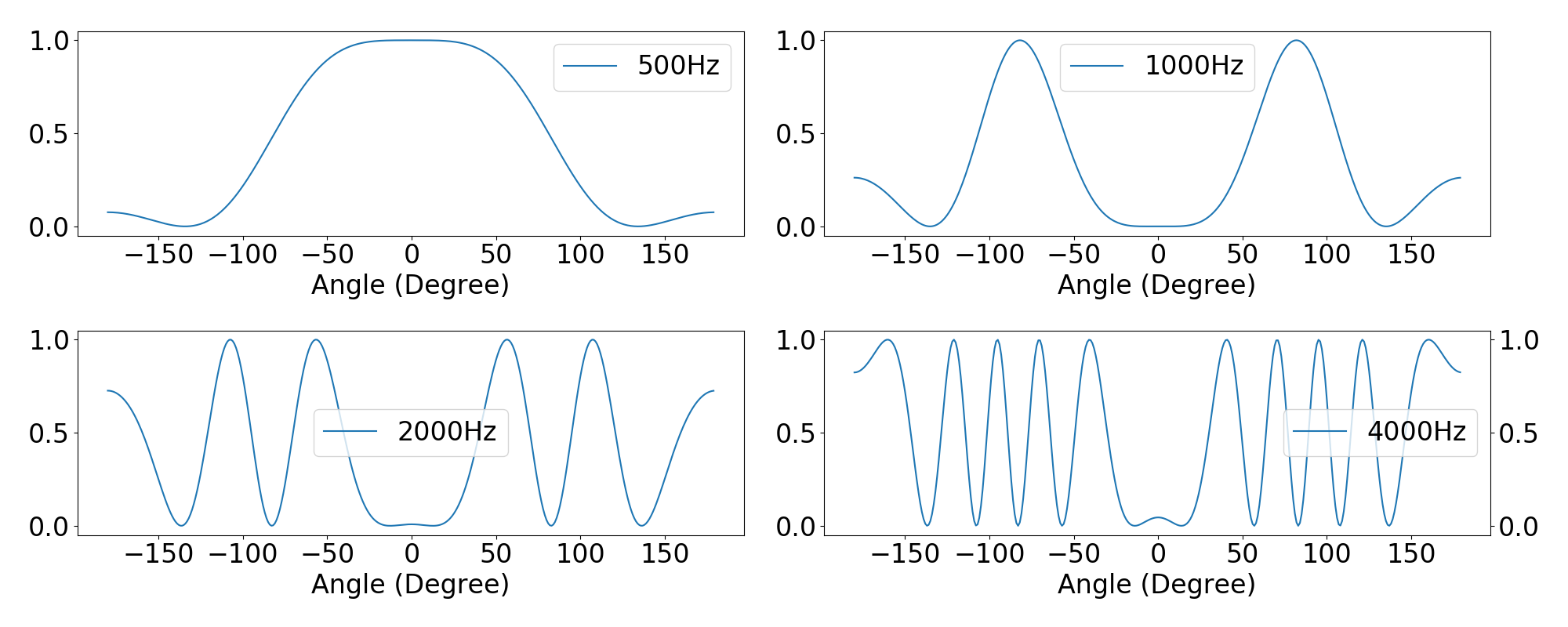}}
  \centerline{(a)Speaker 1}\medskip
\end{minipage}
\begin{minipage}[b]{1.0\linewidth}
  \centering
  \centerline{\includegraphics[width=\linewidth]{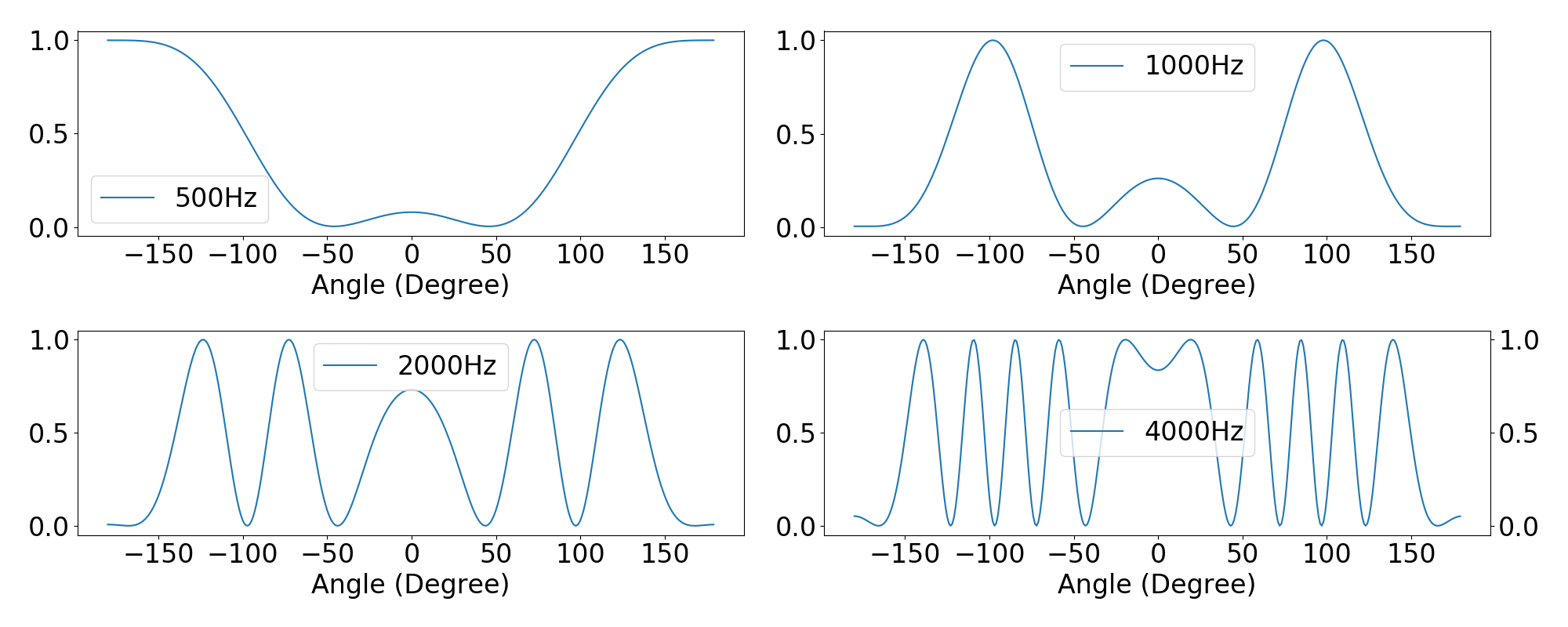}}
  \centerline{(b) Speaker 2}\medskip
\end{minipage}
\vspace{-0.7cm}
\caption{Example of beam patterns of the separated speech.}
\vspace{-0.2cm}
\label{fig:beampattern}
\end{figure}

In order to further explore the neural beamformer's effect,  we show an example of estimated beam pattern \cite{Consolidated-Gannot2017} for the separated sources. Figure~\ref{fig:beampattern} shows the beam pattern of two separated signals at frequencies $\{500\ \text{Hz}, 1000\ \text{Hz},2000\ \text{Hz},4000\ \text{Hz}\}$. The value of the beam at different degrees quantifies the reduction of the speech signals received. As we can see from the figures, the crests and troughs of the beams are different for the two speakers, which shows the neural beamformer is trained properly and can tell the difference between the sources.

\begin{table}[tbh]
\vspace{-0.3cm}

    \caption{Performance in terms of average CER and WER [\%] of the baseline single-speaker end-to-end speech recognition model trained on reverberant (R) single-speaker speech and evaluated on reverberant (R) multi-speaker speech.}
    \label{tab:results-single-asr-reverb}
    \vspace{0.1cm}
    \centering
    \small
    \begin{tabular}{l | c | c}
        \hline \hline
        Model & dev CER (R) & eval CER (R) \\
        \hline
        End-to-End Model (R) & 81.6 & 82.7 \\
        \hline \hline
        Model & dev WER (R) & eval WER (R) \\
        \hline
        End-to-End Model (R) & 103.9 & 104.2 \\
        \hline \hline
    \end{tabular}
\end{table}

\subsection{Evaluation on the spatialized reverberant data}
\label{ssec:result-reverb}
To give a comprehensive analysis of the MIMO-Speech model, we investigated how the model performs in a more realistic case, using the spatialized reverberant wsj1-2mix data. As a comparison, we first trained a normal single-speaker end-to-end speech recognition system. The model is trained with the spatialized reverberant speech from each single speaker. The performance is shown in Table~\ref{tab:results-single-asr-reverb}. For the MIMO-Speech model, the spatialized reverberant wsj1-2mix training dataset was added to the training set for the multi-conditioned training. The results on the speech recognition task are shown in Table~\ref{tab:results-asr-reverb}. The reverberant speech is difficult to recognize as the performance shows severe degradation when we tried to infer the reverberant speech using the anechoic multi-speaker model. The multi-conditioned training can release such degradation, improving the WER by over $60\%$. The results suggest that the proposed MIMO-Speech also has potential for application in complex scenarios. As a complementary experiment, we used Nara-WPE \cite{NaraWPE-Drude2018} to perform speech dereverberation only for the development and evaluation data. The speech recognition results are shown in Table.\ref{tab:results-asr-dereverb} which suggests that the speech dereverberation techniques only in the inference stage can lead to further improvement. Note that the results here are just a preliminary study. The main drawback here is that we did not consider any dereverberation techniques in designing our model.

\begin{table}[th]
    \caption{Performance in terms of average CER and WER [\%] on the spatialized wsj1-2mix corpus of MIMO-Speech trained on either anechoic (A) or reverberant (R) and evaluated on either the anechoic (A) or reverberant (R) evaluation set.}
    \label{tab:results-asr-reverb}
    \vspace{0.1cm}
    \centering
    \small
    \begin{tabular}{l | c | c}
        \hline \hline
        Model & eval CER (A) & eval CER (R) \\
        \hline
        MIMO-Speech (A) & 4.51 & 62.32 \\
        MIMO-Speech (R) & 4.08 & 18.15 \\
        \hline \hline
        Model & eval WER (A) & eval WER (R) \\
        \hline
        MIMO-Speech (A) & 8.62 & 81.30 \\
        MIMO-Speech (R)  & 8.72 & 29.99 \\
        \hline \hline
    \end{tabular}
\end{table}

\begin{table}[th]
    \caption{Performance in terms of average CER and WER [\%] on the spatialized wsj1-2mix corpus of MIMO-Speech trained on either anechoic (A) or reverberant (R) and evaluated on the reverberant data after Nara-WPE dereverberation (D).}
    \label{tab:results-asr-dereverb}
    \vspace{0.1cm}
    \centering
    \small
    \begin{tabular}{l | c | c}
        \hline \hline
        Model & dev CER (D) & eval CER (D) \\
        \hline
        MIMO-Speech (A) & 51.00 & 52.02 \\
        MIMO-Speech (R) & 20.09 & 15.04 \\
        \hline \hline
        Model & dev WER (D) & dev WER (D) \\
        \hline
        MIMO-Speech (A) & 69.08 & 69.42 \\
        MIMO-Speech (R) & 33.09 & 25.28 \\
        \hline \hline
    \end{tabular}
\end{table} 
\section{Conclusion}
\label{sec:conclusion}
In this paper, we have proposed an end-to-end multi-channel multi-speaker speech recognition model called MIMO-Speech. More specifically, the model takes multi-speaker speech recorded by a microphone array as input and outputs text sequences for each speaker. Furthermore, the front-end of the model, involving a neural beamformer, learns to perform speech separation even though no explicit signal reconstruction criterion is used. The main advantage of the proposed approach is that the whole model is differentiable and can be optimized with an ASR loss as target. In order to make the training easier, we utilized single-channel single-speaker speech as well. We also designed an effective curriculum learning strategy to improve the performance. Experiments on a spatialized version of the wsj1-2mix corpus show that the proposed framework has fairly good performance. However, performance on reverberant data still suffers from a large gap against the anechoic case. In future work, we will continue to improve this system by incorporating dereverberation strategies as well as by integrating further the masking and beamforming approaches. %

\section{Acknowledgement}
Wangyou Zhang and Yanmin Qian were supported by the China NSFC projects (No. 61603252 and No. U1736202). %

We are grateful to Xiaofei Wang for sharing his code for beam pattern visualization.

\vfill
\pagebreak
\balance

\bibliographystyle{IEEEtrans_nourl}
\bibliography{refs}

\end{document}